\documentclass[a4paper]{article}

\usepackage[acronym, toc, nonumberlist]{glossaries}
\newacronym{AI}{AI}{artificial intelligence}
\newacronym{ASR}{ASR}{Automatic Speech Recognition}
\newacronym{AU}{AU}{acoustic unit}
\newacronym{AUD}{AUD}{acoustic unit discovery}
\newacronym{BFS}{F-score}{F-score}
\newacronym{BHMMVAE}{BHMMVAE}{Bayesian hidden Markov model variational autoencoder}
\newacronym{CNN}{CNN}{Convolutional Neural Network}
\newacronym{CP}{CP}{cluster purity}
\newacronym{CPC}{CPC}{Contrastive Predictive Coding}
\newacronym{DP}{DP}{Dirichlet process}
\newacronym{ELBO}{ELBO}{evidence lower bound}
\newacronym{EPER}{EPER}{equivalent Phone Error Rate}
\newacronym{EM}{EM}{expectation-maximization}
\newacronym{FB}{FB}{forward-backward}
\newacronym{FHVAE}{FHVAE}{Factorized Hierarchical Variational Autoencoder}
\newacronym{FVAE}{FVAE}{Factorized Variational Autoencoder}
\newacronym{fMLLR}{fMLLR}{Feature Space Maximum Likelihood Linear Regression}
\newacronym{GAP}{GAP}{global average pooling}
\newacronym{GD}{GD}{gradient descent}
\newacronym{GMM}{GMM}{Gaussian mixture model}
\newacronym{GMMVAE}{GMMVAE}{Gaussian mixture model variational autoencoder}
\newacronym{HGMMVAE}{HGMMVAE}{hierarchical gaussian mixture model variational autoencoder}
\newacronym{HMM}{HMM}{hidden Markov model}
\newacronym{HMMVAE}{HMMVAE}{hidden Markov model variational autoencoder}
\newacronym{iid}{i.i.d.}{independent and identically distributed}
\newacronym{KLD}{KLD}{Kullback-Leibler divergence}
\newacronym{KL}{KL}{Kullback-Leibler}
\newacronym{MFCC}{MFCC}{mel frequency cepstral coefficent}
\newacronym{MI}{MI}{mutual information}
\newacronym{MSE}{MSE}{mean squared error}
\newacronym{NMI}{NMI}{normalized mutual information}
\newacronym{NN}{NN}{neural network}
\newacronym{PER}{PER}{phone error rate}
\newacronym{PGM}{PGM}{probabilistic graphical model}
\newacronym{SGD}{SGD}{stochastic gradient descent}
\newacronym{SVAE}{SVAE}{structured VAE}
\newacronym{SVI}{SVI}{stochastic variational inference}
\newacronym{TDP}{DP}{truncated Dirichlet process}
\newacronym{VAE}{VAE}{variational autoencoder}
\newacronym{VC}{VC}{voice conversion}
\newacronym{VI}{VI}{variational inference}
\newacronym{VQVAE}{VQVAE}{Vector Quantized Variational Autoencoder}
\newacronym{VTLN}{VTLN}{Vocal Tract Length Normalization}
\newacronym{WER}{WER}{word error rate}
\newacronym{wrt}{w.r.t.}{with respect to}

\makeglossaries

\usepackage{pgfplots} 
\newlength\fheight 
\newlength\fwidth 
\usepackage{tikz}
\pgfplotsset{compat=1.9}

\usepackage{cite}

\usetikzlibrary{external}

\usepackage{graphicx}
\graphicspath{{./images/}}

\usepackage{nccmath}
\usepackage{bm}
\usepackage{multirow}
\usepackage{balance}
\usepackage{siunitx}

\usepackage{INTERSPEECH2021}

\usepackage[symbol]{footmisc}
\renewcommand{\thefootnote}{\fnsymbol{footnote}}

\title{Voice Conversion Based Speaker Normalization for Acoustic Unit Discovery}
\name{Thomas Glarner$^{*}$,  Janek Ebbers$^{*}$, Reinhold H\"ab-Umbach}
\address{Paderborn University, Germany}
\email{\{glarner,ebbers,haeb\}@nt.upb.de}

\begin{document}

\maketitle
\begin{abstract}
Discovering speaker independent acoustic units purely from spoken input is known to be a hard problem.
In this work we propose an unsupervised speaker normalization technique prior to unit discovery. It is based on
separating speaker related  from content induced variations in a speech signal with an  adversarial contrastive predictive coding approach. This technique does neither require transcribed speech nor speaker labels, and, furthermore, can be trained in a  multilingual fashion, thus achieving speaker normalization even if only few unlabeled data is available from the target language. The speaker normalization is done by mapping all utterances to a medoid style which is representative for the whole database.
We demonstrate the effectiveness of the approach by conducting acoustic unit discovery with a hidden Markov model variational autoencoder noting, however, that the proposed speaker normalization can serve as a front end to any unit discovery system. Experiments on English, Yoruba and Mboshi show improvements compared to using non-normalized input.
\end{abstract}

\let\thefootnote\relax\footnotetext{$^{*}$First two authors contributed equally to this work.}

\section{Introduction}\label{sec:intro}
Modern high performance speech recognition systems require thousands of hours of transcribed speech to train an acoustic model. This stands in sharp contrast to how infants acquire a language. They show an innate ability to infer linguistic structure from the speech signal alone, long before they learn to read and write \cite{Dupoux2018}. The research field of unsupervised speech learning is concerned with the task to endow machines with a similar ability \cite{Ondel2021}. Learning acoustic and language models from spoken input has other applications next to providing a computational model of infant language acquisition. It can be an important tool for linguists in their endeavor to document endangered languages, many of them having no written form. Furthermore, it can help to widen the range of languages for which automatic language processing tools, such as \gls{ASR} systems, can be built, because the constraint of requiring many hours of transcribed speech for their training can be relaxed.

In this work we are concerned with unsupervised \gls{AUD}. With \glspl{AU} we denote the basic building blocks of speech. These are recurring segments of high similarity which we wish to cluster into syntactic classes. We call them \glspl{AU} rather than phones, since they are defined upon acoustic similarity and may not necessarily correspond to linguistically defined units.

We wish those automatically learnt units to capture content information and be independent of speaker characteristics. However, in the considered unsupervised scenario this is known to be a tough goal~\cite{jansen2013clsp, Jansen2013, Walter2013}.

This work proposes a novel approach to speaker normalization: Recent advances in generative and predictive modeling like auxiliary adversarial classifiers and \gls{CPC}~\cite{vandenoord2018cpc} have been utilized to build \gls{VC} systems with high performance. The core idea of this approach is to disentangle the speech input into content and style embeddings. Here, the content embedding covers short term variations in the utterance which are strongly influenced by the phonetic content, while the style embedding covers the information of factors which show only slight variations during the utterance.

\gls{VC} can be achieved by keeping the content embedding while exchanging the style embedding with one extracted from an utterance spoken by the target speaker and reconstructing an audio signal from these embeddings through a generative model.
Expanding this approach, speaker normalization is possible by converting all utterances in the database to the same target style, since the style embedding is mostly influenced by the speaker characteristics.

We design the proposed speaker normalization as a front end of an \gls{AUD} system, such that it can be combined with any approach to \gls{AUD}. In the experiments we employ our earlier proposed \gls{HMMVAE}~\cite{ebbers2017hmmvae}, which has a \gls{VAE} structure with a neural encoder and decoder. Unlike conventional \glspl{VAE},  \glspl{HMM} are employed to model the latent space variables, one per \gls{AU}, to capture the temporal structure of \glspl{AU}. The \gls{HMMVAE} has been shown in prior work to outperform HMM-GMM based \gls{AUD} systems~\cite{ebbers2017hmmvae, glarner2018fbhmmvae}.

The remainder of this work is organized as follows. Section~\ref{sec:rel_work} discusses related work. In Section~\ref{sec:prop_sys}, the proposed system is explained in detail, with the Adversarial CPC based \gls{VC} system covered in Section \ref{ssec:ACPC_VC}, and the \gls{HMMVAE} discussed in \ref{ssec:HMMVAE}.
In Section~\ref{sec:ex_setup}, the experiment setup is explained, covering feature extraction in Section~\ref{ssec:features} and the training details in Section~\ref{ssec:train}.
Section~\ref{ssec:measures} introduces the performance metrics used to evaluate the system.
The results are presented and discussed in Section~\ref{sec:results}.
Finally, conclusions are drawn in Section \ref{sec:conclusions}.

\begin{figure*}[t!]
  \vspace{-1.5em}
  \begin{minipage}{.5\textwidth}
    \centering
    \includegraphics[width=.85\linewidth]{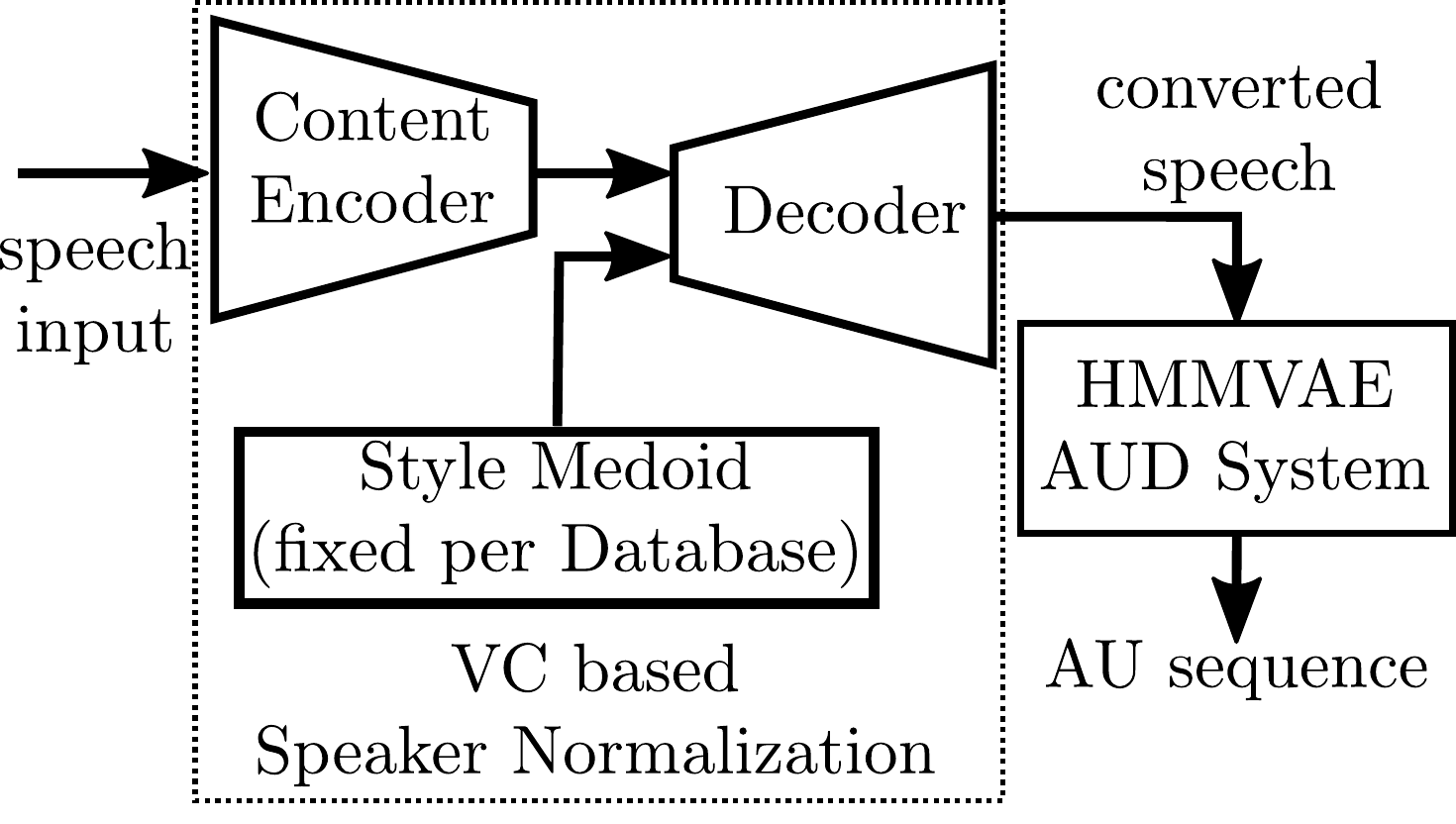}
    \captionof{figure}{\gls{AUD} with Speaker Normalization}
    \label{fig:aud_system}
  \end{minipage}%
  \begin{minipage}{.5\textwidth}
    \centering
    \includegraphics[width=.85\linewidth]{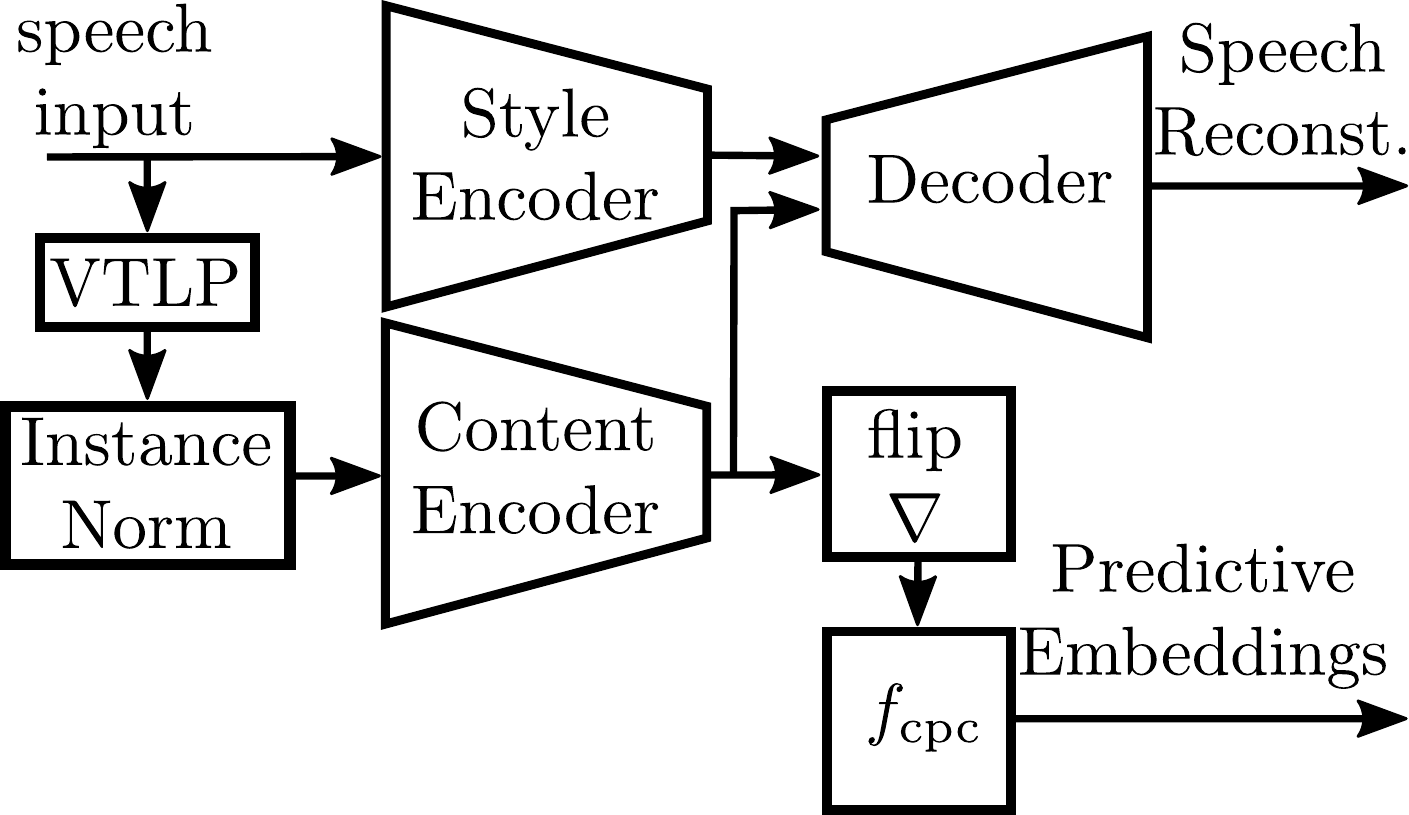}
    \captionof{figure}{\gls{VC} system training}
    \label{fig:train_vc}
  \end{minipage}%
  \vspace{-1.5em}
\end{figure*}

\vspace{-0.5em}
\section{Related Work}\label{sec:rel_work}
A classic approach to tackle the problem of speaker variablity in Automatic Speech Recognition is \gls{VTLN}~\cite{acero1990acousticaland,eide1996vtln,lee1996snwarping}, which has also been applied in low-resource settings, e.g., for spoken term detection~\cite{madhavi2019vtlnforsd}.
Furthermore, \gls{fMLLR} is often used, even in low-resource setups~\cite{heck2017dpgmm}.
However, a significant amount of training data is needed per speaker and the speaker labels need to be known.

In the Zero Resource Speech challenge 2017~\cite{dunbar2017zerospeech}, Track 1 was geared towards improving \gls{AUD} results with speaker adaptation techniques and thus provided speaker labels.
In contrast to this, our approach to speaker normalization is completely unsupervised and does not require speaker labels.

The Zero Resource Speech challenge 2019~\cite{dunbar2019zerospeech} introduced a setup where \gls{AUD} systems are used to provide the content input of a voice synthesis system.
Leading systems of the challenge~\cite{tjandra2019zerospeech_vqvae, feng2019disentangle_zerospeech} are based on the \gls{VQVAE} framework~\cite{vandenoord2017vqvae, chorowski2019vqvae_rep} or the \gls{FHVAE}~\cite{hsu2017fhvae}.
They also employ the idea of disentangling content from speaker characteristics, to extract the content information from the input.
Our approach is reversed as we aim at extracting and removing the speaker information.
Furthermore, we more actively pursue disentanglement by placing an adversarial CPC loss on the content representation.

In~\cite{yusuf2020hierarchical}, a multilingual \gls{AUD} system is constructed which defines a subspace of \glspl{AU} which is learned in a supervised way from multilingual data in an attempt to capture the commonalities on what an \gls{AU} is across different languages.
They aim at providing a better prior for the \gls{AU} learning, while we are concerned with removing speaker dependence.
We thus conjecture that the two approaches are complementary and possibly could be combined.

\vspace{-0.5em}
\section{System Setup}\label{sec:prop_sys}
The proposed system is shown in Figure~\ref{fig:aud_system}.
It consists of an adversarial \gls{CPC} based \gls{VC} system~\cite{ebbers2020contrastive} for speaker normalization and a subsequent \gls{HMMVAE}~\cite{ebbers2017hmmvae} to perform the \gls{AUD}.

\subsection{Adversarial CPC Based Voice Conversion}\label{ssec:ACPC_VC}
Unsupervised disentanglement of speaker and content induced variations in speech signals has attracted increased popularity in recent years.
One reason is that it shows a way to exploit unlabeled data to, e.g., obtain a content representation, which is invariant to the speaker and, thus, may enable improved performance in a downstream task.
With most approaches employing some kind of auto encoding they further allow to perform voice conversion by decoding the content representation of an utterance with the speaker representation of some other utterance.

As we believe that the speaker induced variations in the speech signals are impairing the performance of \gls{AUD} systems, we propose to perform a speaker normalization beforehand.
Here, one could either apply an \gls{AUD} system to the speaker invariant latent content representation or one could convert each utterance to the same speaker before inputting it to the \gls{AUD} system.
In this paper, we decide for the latter approach as \gls{AUD} systems can then be applied to conversions in just the same way as they could be applied to non-converted signals.

For the voice conversion we here employ a \gls{FVAE} along with adversarial \gls{CPC} as proposed in \cite{ebbers2020contrastive}, which has shown to yield a well-balanced trade-off between linguistic content preservation and speaker invariance.
The \gls{FVAE} employs two convolutional encoders, namely, a content encoder outputting a series of content embeddings ${\mathbf{C}=(\mathbf{c}_{1}, \dots, \mathbf{c}_{M})}$ and a style encoder with subsequent \gls{GAP} yielding an utterance-level style embedding $\mathbf{s}$, see Fig.~\ref{fig:train_vc}.
Note that the encoders use temporal pooling which is why $M<T$.
The encoders are trained jointly with a decoder $\hat{\mathbf{Y}}=f_\text{dec}(\mathbf{C}, \mathbf{s})$ to allow reconstruction of the speech signal $\mathbf{Y}$ by employing a reconstruction loss ${L_\text{rec}=\frac{1}{T}||\hat{\mathbf{Y}}-\mathbf{Y}||^2_2}$.
Note that  the encoder also follows the \gls{VAE} framework and provides the statistics of a variational posterior $q(\mathbf{c}_m)$ rather than $\mathbf{c}_m$ directly, which, however, is neglected in our notation for the sake of clarity.
Nonetheless, it contributes a \gls{KL} regularization loss ${L_\text{kld} = \frac{1}{M}\mathrm{D}_\text{KL}(q(\mathbf{c}_m; \bm\phi)\|p(\mathbf{c}_m))}$ with $p(\mathbf{c}_m)$ being a standard normal prior.
To foster disentanglement, adversarial \gls{CPC} is employed to prevent the content encoder from encoding style information.

\gls{CPC} is an increasingly popular self-supervised learning approach, allowing to extract mutual information of different parts of the same signal by employing a contrastive loss.
Here, however, it is to identify static information in the content embeddings and repel it.
For that purpose, a CPC encoder $\mathbf{h}_t = f_\text{CPC}(\mathbf{C}'_t)$, with $\mathbf{C}'_t$ denoting the receptive field of $f_\text{CPC}$ at frame $t$, is trained to extract mutual information from $\mathbf{C}'_{t-\tau}$ and $\mathbf{C}'_{t}$ by employing a contrastive loss
\begin{align*}
\label{eq:cpc}
L_\text{cpc} = -\frac{1}{T-\tau}\sum_{t=\tau + 1}^T\frac{\exp(\mathbf{h}_t^\mathrm{T}\mathbf{h}_{t-\tau})}{\sum\limits_{\widetilde{\mathbf{h}}_t\in\mathcal{B}_t} \exp(\widetilde{\mathbf{h}}_t^\mathrm{T}\mathbf{h}_{t-\tau})}\,\,
\end{align*}
where $\tau$ denotes a lookahead shift and $\mathcal{B}_t$ presents a set of candidate embeddings $\tilde{\mathbf{h}}_t$.
By choosing the lookahead shift to correspond to one second, the encoded mutual information that may be extracted represents static attributes such as style information.
While the CPC encoder is trained to minimize $L_\text{cpc}$, the content encoder is trained to maximize it, i.e. content and \gls{CPC} encoder are operating adversarially here, which encourages the \gls{FVAE} to not encode style information into the content embeddings but only into the style embedding.
Note that computation of $L_\text{cpc}$ does not require any labels but solely relies on self-supervision.
The overall \gls{FVAE} loss is given as $$L_\text{fvae}=L_\text{rec}+\beta L_\text{kld}-\lambda L_\text{cpc},$$ where we choose $\beta=0.01$ and $\lambda = 1$ following \cite{ebbers2020contrastive}.

After training, the \gls{FVAE} is used for speaker normalization, by converting all utterances of a certain database to the same target style.
Ideally, the target style is supposed to be an average style which is representative for the respective database.
However, to avoid introducing unnecessary artifacts due to a nonexistent target speaker, we here use the style medoid instead.
Firstly, the style embedding vector of each utterance in the respective database is extracted.
Then, the style embedding vector which has the minimal mean Euclidean distance to all the style embeddings in the database,  is chosen as our target style, which all utterances are converted to.
The converted utterances are then forwarded to the \gls{AUD} system, which here is implemented by an \gls{HMMVAE} as explained in the following.

\subsection{HMMVAE}\label{ssec:HMMVAE}
The \gls{HMMVAE} is based on a \gls{VAE}~\cite{kingma2013auto}, which is a very well-known generative model.
Here, we limit ourselves to normal distributions, with  $p(\mathbf{y}|\mathbf{x}) = \mathcal{N}(\mathbf{y}; f(\mathbf{x}; {\theta}), \sigma^2 \cdot \mathbf{I})$ as the conditional distribution of the observation $\mathbf{x}$ given the latent code $\mathbf{x}$, and a decoder \gls{NN} \gls{NN} $f(\cdot; \bm{\theta})$.
The variational~\cite{blei2017variational} posterior is given by $q(\mathbf{x}_n, \bm\phi) = \mathcal{N}(\mathbf{x}_n; \bm{\mu}_\mathbf{x}(\mathbf{y}_n; \bm\phi), \mathrm{diag}\{\bm{\sigma}_{\mathbf{x}}^2(\mathbf{y}_n; \bm\phi)\})$,
with a decoder \gls{NN} to extract the mean vector $\bm{\mu}_\mathbf{x}(\mathbf{y}_n; \bm\phi)$ and the log-variances $\log\bm{\sigma}_{\mathbf{x}}^2(\mathbf{y}_n; \bm\phi)$.

Parameters are learned by minimizing the negative \gls{ELBO}
\vspace{-0.5em}
\begin{align*}
  &- \log p(\mathbf{y}_n) \le -\mathrm{ELBO}^\text{(VAE)} \\
    &= \frac{\mathbb{E}_{q(\mathbf{x}_n; \bm\phi)}\| \mathbf{y}_n - f(\mathbf{x}_n; \bm\theta)\|^2}{2 \sigma^2} + \mathrm{D}_\text{KL}(q(\mathbf{x}_n; \bm\phi)\|p(\mathbf{x}_n)) + \mathrm{C.}
\end{align*}
over a dataset $\mathcal{Y} = \{\mathbf{y}_1, \dots, \mathbf{y}_N\}$ of $N$ observations.

The intractability of the first term, called the reconstruction loss, is side-stepped by sampling from the posterior.
Note that with a fixed observation variance $\sigma^2$, the model can be interpreted as a $\beta$-VAE~\cite{higgins2016betaVAE}.

In the vanilla \gls{VAE}, a standard normal distribution is assumed for the latent prior $p(\mathbf{x}_n)$.
For the \gls{HMMVAE}, which is an instance of the \gls{SVAE}~\cite{johnson2016composing}, this is replaced by an \gls{HMM}.
Here, $\mathbf{Y} {=} (\mathbf{y}_{1}, \dots, \mathbf{y}_{T})$ denotes the observations from a single utterance of length $T$.
The corresponding sequence of latent codes, which are emitted by the \gls{HMM}, is $\mathbf{X} {=} (\mathbf{x}_{1}, \dots, \mathbf{x}_{T})$.
The corresponding latent state sequence is $\mathbf{Z} {=} (\mathbf{z}_{1}, \dots, \mathbf{z}_{T})$,
which consists of a one-hot vector $\mathbf{z}_t = [z_{t,1}, \dots, z_{t,N_\text{S}}]^\mathrm{T}$ for each time step $t$ with the number of states $N_\text{S}$ as length.
The state-dependent emission densities are $\mathcal{N}(\mathbf{x}_t; \bm{\mu}_k, \bm\Sigma_k)$, and the state transition matrix is given by $\mathbf{A}$.
The full \gls{HMM} parameter set to be learned is thus $\bm\Omega {=} (\mathbf{A},  \bm\mu_1, \bm{\Sigma}_1, \dots, \bm{\mu}_{N_\text{S}}, \bm{\Sigma}_{N_\text{S}})$,
and the latent prior becomes $p(\mathbf{X}, \mathbf{Z}; \bm\Omega)$.
Constraints on the parameters are fulfilled by proper parametrization as in~\cite{ebbers2017hmmvae}.
The variational posterior needs to be expanded to $q(\mathbf{X}, \mathbf{Z}) {=} q(\mathbf{Z}) \prod_{t=1}^{T}q(\mathbf{x}_t; \bm\phi)$ to include the state sequence.

For the \gls{HMMVAE}, the \gls{ELBO} thus becomes
\vspace{-0.5em}
\begin{align*}
  \mathrm{ELBO}^{\text{(HMMVAE)}}
  = \mathbb{E}_{q(\mathbf{X},\mathbf{Z})} \left[\log \frac{p(\mathbf{Y}| \mathbf{X}) p(\mathbf{X}, \mathbf{Z}; \bm\theta, \bm\Omega)}{q(\mathbf{X}, \mathbf{Z}; \bm{\phi}, \bm\Omega)} \right].
\end{align*}
Note that this expression can be further decomposed due to the mean-field approximation and the fact that the observation $\mathbf{y}_t$ depends only on $\mathbf{x}_t$.
Consequently, the same reconstruction loss term appears as in the vanilla \gls{VAE}.

The state posterior $q(z_t)$ and joint posterior $q(z_{t{-}1}, z_t)$, which are needed for inference, can be efficiently approximated by Viterbi~\cite{forney1973viterbi} estimates $\hat{\mathbf{z}}^{\mathcal{V}}_{t}$, i.e.
$q(\mathbf{z}_t) \rightarrow \delta(\mathbf{z}_t {=} \hat{\mathbf{z}}^{\mathcal{V}}_{t})$, and
$q(\mathbf{z}_{t{-}1}, \mathbf{z}_t) \rightarrow \delta(\mathbf{z}_{t-1} {=} \hat{\mathbf{z}}^{\mathcal{V}}_{t-1}) {\cdot} \delta(\mathbf{z}_t {=} \hat{\mathbf{z}}^{\mathcal{V}}_{t})$ , respectively.

In the case of \gls{AUD} each \gls{AU} is modeled by three states, giving $N_\text{S} = 3\cdot N_\text{U}$, with $N_\text{U}$ being the number of \glspl{AU}.

\vspace{-0.5em}
\section{Experiment Setup}\label{sec:ex_setup}
To evaluate the performance impact of our speaker normalization approach on the \gls{AUD} task, experiments on three languages were conducted.

\subsection{Speech Databases}\label{ssec:speech_db}
The \gls{VC} system is trained on all available data at once, i.e., across the languages available.
This is plausible since the \gls{VC} system attempts to capture general properties of the human vocal tract which are mostly the same for all speakers regardless of the target language.
Furthermore, the voice conversion system needs a certain amount of training data to avoid overfitting the adversarial classifier as well as a certain amount of speaker diversity to avoid having the same two speakers in the same batch too often as this harms contrastive learning.
The first language, English, serves two purposes: As a control language and as a resource for large amounts of data during training of the \gls{VC} system.
Therefore, we use two databases: The first is TIMIT~\cite{garofolo1993timit}, where the sa utterances are left out, leading to \SI{3.6}{\hour} of recordings with 4288 utterances by 536 speakers. Second, LibriSpeech~\cite{panayotov2015librispeech} is used solely to train the \gls{VC} system, with the subsets \texttt{train-clean-100} and \texttt{train-clean-360}, adding up to \SI{464.2}{\hour} of speech with  132553 utterances by 1172 speakers.

Furthermore, the West African language Yoruba~\cite{gutkin2020yoruba}, including \SI{4}{\hour} of recordings with 3583 utterances by 36 speakers, and the Central African language Mboshi~\cite{godard2017mbochi}, containing \SI{4.4}{\hour} of recordings with 5130 utterances by 3 speakers, are used to demonstrate the performance on  low-resource languages. Available transcriptions are used for evaluation only. In the case of TIMIT, the proper phone-level transcriptions are used. For Yoruba and Mboshi, forced alignments were provided by the authors of~\cite{yusuf2020hierarchical}.
The databases for \gls{AUD} are deliberately chosen as in~\cite{yusuf2020hierarchical} and \cite{Ondel2021} to provide comparability.

\subsection{Feature Extraction}\label{ssec:features}
In this work, a log-mel feature representation is chosen, which consists of taking the short-time Fourier transform of the input audio, conversion to the power spectrum domain, applying a mel filterbank  and taking the logarithm of the resulting mel power spectrum.
All data is processed with a sampling rate of \SI{16}{\kilo\hertz}.
For the \gls{VC} system, the parameters are: A Blackman window with a window length of 400 samples, a window shift of 160 samples, an FFT size of 512 samples and 80 filters in the mel filterbank. These parameters are taken from~\cite{ebbers2020contrastive}.
Each of the mel-bands is normalized to zero mean and unit variance, where normalization statistics are computed using the complete training set.

For the HMMVAE, we use 40 filters in the mel filterbank and deltas and delta-deltas of the log-mel features are added.
Furthermore, each of the feature maps is normalized to zero mean and unit variance for each input signal individually.

\subsection{Training}\label{ssec:train}
Both models are trained using Adam~\cite{kingma2014adam}.
The \gls{VC} system is trained in a multilingual manner on all datasets of all language databases.
Since the \gls{CPC} part uses other examples from the same batch as counter examples, it is ensured that each batch only contains utterances from a single language to avoid easy predictability for the adversarial classifier due to language characteristics.
The batch size is chosen to be 16.

While the \gls{VC} system is trained on multilingual data, the \gls{HMMVAE} has to be trained language spefic.
Its input audio is either taken directly from the database in the control experiments or from the output of the \gls{VC} system.
For all languages, the number of \glspl{AU} $N_\text{U}$ is set to 80.
Optimization is done with the ADAM optimizer~\cite{kingma2014adam} with a learning rate of $0.001$.
Similar to~\cite{ebbers2017hmmvae}, a pre-training scheme is used to initialize the parameters, where the \gls{HMMVAE} is trained in a pseudo-supervised manner on random alignments for 2000 iterations.
The actual HMMVAE training is carried out for 20000 iterations since this has shown to be enough for the system to converge.

\subsection{Performance Metrics}\label{ssec:measures}
The performance is evaluated using three different metrics. For each metric, higher values are better.
Firstly, the (symmetric) \gls{NMI} is used as defined in~\cite{yusuf2020hierarchical}:
$\mathrm{NMI} = \SI{200}{\percent} \frac{\mathrm{I}(U;P)}{\mathrm{H}(U) + \mathrm{H}(P)}$, where $U$ are the extracted \glspl{AU}, $P$ are the reference phones, $\mathrm{I}(\cdot;\cdot)$ is the mutual information between the label sets and $\mathrm{H}(\cdot)$ is the entropy.
The calculation is based on a frame-wise comparison of all proposed and reference transcriptions, calculating a confusion matrix and estimating the joint probability distribution between the two label sets.
In comparison to the older (asymmetric) \gls{NMI} definition, using too many \glspl{AU} induces a penalty, which leads to a better comparability across different choices for the number of \glspl{AU}.
Secondly the \gls{CP} of each \gls{AU} cluster is calculated from the confusion matrix. These first two metrics assess the cluster performance.

As the third measure, we calculate the phone boundary \gls{BFS}, which assesses the agreement of the discovered \gls{AU} boundaries with the phone boundaries provided by the database (TIMIT) or by forced alignment (Yoruba, Mboshi), using a collar of $\pm \SI{20}{\milli\second}$ as in~\cite{yusuf2020hierarchical}. This metric measures the segmentation performance.

\section{Results}\label{sec:results}

\begin{table}[t!]
  \caption{Comparison of \gls{AUD} results}
  \label{tbl:results}
  \setlength\tabcolsep{5pt}
  \centering
  \resizebox{\linewidth}{!}{
  \begin{tabular}{lllccc}
    \toprule
    language & model                               & input          & \acrshort{NMI}     & \acrshort{CP}       & \acrshort{BFS}      \\
    \midrule
    English  & HMM~\cite{yusuf2020hierarchical}    & clean          & 35.91 {\scriptsize${\pm}$0.27}  & ---                 & 63.86 {\scriptsize${\pm}$0.34}  \\
             & H-SHMM~\cite{yusuf2020hierarchical} & clean          & 40.04 {\scriptsize${\pm}$0.51}  & ---                 & 76.60 {\scriptsize${\pm}$0.54}  \\
             & HMMVAE                              & clean          & 40.44 {\scriptsize${\pm}$0.18}      & 39.00 {\scriptsize${\pm}$0.57}       & 75.18 {\scriptsize${\pm}$0.20}       \\
             & HMMVAE                              & rec            & 41.14 {\scriptsize${\pm}$0.34}      & 41.01 {\scriptsize${\pm}$0.36}       & 76.03 {\scriptsize${\pm}$0.64}       \\
             & HMMVAE                              & vc           & 42.85 {\scriptsize${\pm}$0.28}      & 42.09 {\scriptsize${\pm}$0.72}       & 73.88 {\scriptsize${\pm}$0.62}       \\
             \midrule
    Mboshi   & HMM~\cite{yusuf2020hierarchical}    & clean          & 35.85 {\scriptsize${\pm}$0.62}  &  ---                & 47.92 {\scriptsize${\pm}$1.56}   \\
             & H-SHMM~\cite{yusuf2020hierarchical} & clean          & 41.07 {\scriptsize${\pm}$1.09}  &  ---                & 59.15 {\scriptsize${\pm}$1.51}   \\
             & HMMVAE                              & clean          & 35.87 {\scriptsize${\pm}$0.59}      & 58.39 {\scriptsize${\pm}$0.39}       & 53.54 {\scriptsize${\pm}$2.34}       \\
             & HMMVAE                              & rec            & 36.33 {\scriptsize${\pm}$0.43}      & 58.84 {\scriptsize${\pm}$0.22}       & 53.82 {\scriptsize${\pm}$1.33}       \\
             & HMMVAE                              & vc           & 38.13 {\scriptsize${\pm}$0.48}      & 58.58 {\scriptsize${\pm}$0.22}       & 53.33 {\scriptsize${\pm}$1.03}       \\
             \midrule
    Yoruba   & HMM~\cite{yusuf2020hierarchical}    & clean          & 36.38 {\scriptsize${\pm}$0.22}  &  ---                & 54.47 {\scriptsize${\pm}$0.64}   \\
             & H-SHMM~\cite{yusuf2020hierarchical} & clean          & 40.06 {\scriptsize${\pm}$0.11}  &  ---                & 66.95 {\scriptsize${\pm}$0.36}   \\
             & HMMVAE                              & clean          & 36.64 {\scriptsize${\pm}$0.36}      & 48.02 {\scriptsize${\pm}$0.28}       & 55.26 {\scriptsize${\pm}$0.93}       \\
             & HMMVAE                              & rec           & 38.14 {\scriptsize${\pm}$0.29}      & 50.16 {\scriptsize${\pm}$0.32}       & 55.97 {\scriptsize${\pm}$0.35}       \\
             & HMMVAE                              & vc           & 38.72 {\scriptsize${\pm}$0.32}      & 50.40 {\scriptsize${\pm}$0.35}       & 54.41 {\scriptsize${\pm}$0.81}       \\
             \midrule
             \bottomrule
  \end{tabular}
  }
  \vspace{-.5em}
\end{table}

We conducted three experiments per language:
\gls{AUD} without voice conversion, labeled as \texttt{clean},
\gls{AUD} after reconstruction, labeled \texttt{rec} where source style and target style were the same,
and \gls{AUD} after speaker normalization by \gls{VC} to the style medoid of the respective database, labeled \texttt{vc}.
Each system is trained 5 times with different random weight initialization and we report mean and variance of the metrics across the different runs. The results are presented in Table~\ref{tbl:results}.

It can be seen that for English a significant gain of \SI{2.41}{\%} and \SI{3.09}{\%} can be achieved for \gls{NMI} and \gls{CP}, respectively, by using voice conversion based speaker normalization even allowing to outperform the H-SHMM model.
While the \gls{NMI} and \gls{CP} show that the cluster quality can be improved, the \gls{BFS} shows that the segmentation performance is reduced with voice conversion based speaker normalization.
Note, however, that the proposed speaker normalization is intended to better find similar units rather than finding more precise boundaries.
Indeed, it makes intuitively sense that with voice conversion the segment boundaries may deviate from the original boundaries more frequently, given that the conversion may also slightly shift phoneme boundaries.
That this might be the case, is also supported by the fact that the \gls{BFS} does not decrease when using reconstruction instead of voice conversion.

For Yoruba and Mboshi speaker normalization also brings a gain in cluster quality, which, however, is more modest for these languages and the H-SHMM model from~\cite{yusuf2020hierarchical} shows on average a better clustering performance than the \gls{HMMVAE}.
The lower gain for these languages probably results from the fact that the \gls{VC} system has primarily been trained on English data.
Nonetheless, the results show that improvements can indeed be achieved through normalization by multilingual voice conversion.
Furthermore, with the availability of larger speech corpora for these languages -- even completely untranscribed and unannotated -- which would allow the language to have a larger impact on \gls{VC} training, higher gains may be achieved.
Finally, since our proposed speaker normalization is rather generic it may be also combined with the H-SHMM in future experiments.

Generally it can be observed that experiments for Mboshi and Yoruba show both a higher \gls{CP} than experiments on English while simultaneously giving a lower \gls{NMI}.
Note that, while the \gls{NMI} and the \gls{CP} both measure the clustering performance, the \gls{NMI} has a higher weighting of infrequently appearing \glspl{AU}, which might explain this discrepancy.
Additionally, both African languages have a larger phone inventory and, consequently, higher phone entropy $\mathrm{H}(P)$ than English.

Interestingly, results with reconstructed signals show on average better performance in all metrics for all languages compared to the experiments with clean inputs.
A possible explanation might be that the voice conversion removes some irrelevant factors on the audio input and thus helps for the \gls{AUD} results to be more robust.


\section{Conclusions}\label{sec:conclusions}
In this work, we have proposed an approach to improve \gls{AUD} with a voice conversion based speaker normalization system
which can be trained in a completely unsupervised manner without any transcriptions and annotations.
We have shown that significant performance improvements are achieved if enough training data in the respective language is available for the \gls{VC} system.
Furthermore, some improvements in the clustering performance can even be achieved even when the training data stems mostly from another language.
Finally, the proposed speaker normalization system is generic and can be combined with any \gls{AUD} system working on audio input.

\section{Acknowledgements}
This work was in part funded by the Deutsche Forschungsgemeinschaft (DFG, German Research Foundation) under project No. 282835863.

\balance
\bibliographystyle{IEEEtran}
\bibliography{vnaud}

\end{document}